\documentclass[sigconf]{acmart}

\AtBeginDocument{%
  \providecommand\BibTeX{{%
    \normalfont B\kern-0.5em{\scshape i\kern-0.25em b}\kern-0.8em\TeX}}}

\copyrightyear{2022} 
\acmYear{2022} 
\setcopyright{rightsretained} 
\acmConference[FPGA '22]{Proceedings of the 2022 ACM/SIGDA International Symposium on Field-Programmable Gate Arrays}{February 27-March 1, 2022}{Virtual Event, CA, USA}
\acmBooktitle{Proceedings of the 2022 ACM/SIGDA International Symposium on Field-Programmable Gate Arrays (FPGA '22), February 27-March 1, 2022, Virtual Event, CA, USA}
\acmDOI{10.1145/3490422.3502357}
\acmISBN{978-1-4503-9149-8/22/02}

\usepackage[normalem]{ulem}

\usepackage{amsmath,amsfonts}
\usepackage{algorithm}
\usepackage{graphicx}
\usepackage{textcomp}
\usepackage{xcolor}
\usepackage{xspace}

\usepackage{algcompatible}

\algblockdefx{FORALLP}{ENDFAP}[1]%
  {\textbf{for all }#1 \textbf{do in parallel}}%
  {\textbf{end for}}

\usepackage{threeparttable}

\usepackage{newfloat}
\DeclareFloatingEnvironment[
    fileext=los,
    listname={List of TableA},
    name=Table A,
    placement=tbhp,
]{tableA}

\newcommand{\sextans}{\textsc{Sextans}\xspace}
\newcommand{\sextansFPGA}{\textsc{Sextans}\xspace}
\newcommand{\sextansASIC}{\textsc{Sextans-P}\xspace}

\newcommand{\authorsqwe}{Linghao Song, 
Yuze Chi,
Atefeh Sohrabizadeh,
Young-kyu Choi,
Jason Lau, and
Jason Cong}

\usepackage{tikz}
\newcommand{\ballnumber}[1]{\tikz[baseline=(myanchor.base)] \node[circle,fill=.,inner sep=1pt] (myanchor) {\color{-.}\bfseries\footnotesize #1};}

\ccsdesc[500]{Hardware~Hardware accelerators}
\ccsdesc[500]{Computer systems organization~Reconfigurable computing}

\keywords{Accelerator, SpMM, Hardware Flexibility, High Bandwidth Memory.}

\begin{document}

\title[\sextans: A Streaming Accelerator for General-Purpose Sparse-Matrix Dense-Matrix Multiplication]{
\sextans: A Streaming Accelerator for \\
General-Purpose Sparse-Matrix Dense-Matrix Multiplication}

\date{}

\author[Linghao Song, 
Yuze Chi,
Atefeh Sohrabizadeh,
Young-kyu Choi,
Jason Lau, and
Jason Cong]{
Linghao Song, 
Yuze Chi,
Atefeh Sohrabizadeh,
Young-kyu Choi$^\dagger$,
Jason Lau, and
Jason Cong
}

\affiliation{%
  \institution{University of California,  Los Angeles \hspace{0.25cm} $^\dagger$Inha University}
  \country{}
}
\email{{linghaosong,chiyuze,atefehsz,lau,cong}@cs.ucla.edu,  ykc@inha.ac.kr}


\fancyhead{}

\begin{abstract}
Sparse-Matrix Dense-Matrix multiplication (SpMM) is the key operator for a wide range of applications including scientific computing, graph processing, and deep learning. Architecting accelerators for SpMM is faced with three challenges -- (1) the random memory accessing and unbalanced load in processing because of random distribution of elements in sparse matrices, (2) inefficient data handling of the large matrices which can not be fit on-chip, and (3) a non-general-purpose accelerator design where one accelerator can only process a fixed-size problem.

In this paper, we present \sextans, an accelerator for general-purpose SpMM processing. \sextans accelerator features (1) fast random access using on-chip memory, (2) streaming access to off-chip large matrices, (3) PE-aware non-zero scheduling for balanced workload with an II=1 pipeline, and (4) hardware flexibility to enable 
prototyping the hardware once
to support SpMMs of different size as a general-purpose
accelerator. We leverage high bandwidth memory (HBM) for the efficient accessing of both sparse and dense matrices. In the evaluation, we present
an FPGA prototype \sextansFPGA which is executable on a Xilinx U280 HBM FPGA board and a projected prototype \sextansASIC with higher bandwidth competitive to V100 and more frequency optimization.
We conduct a comprehensive evaluation on 1,400 SpMMs on a wide range of sparse matrices including 50 matrices from SNAP and 150 from SuiteSparse. We compare \sextans with NVIDIA K80 and V100 GPUs. \sextansFPGA achieves a 2.50x geomean speedup over K80 GPU and \sextansASIC achieves a 1.14x geomean speedup over V100 GPU (4.94x over K80).
The code is available at \url{https://github.com/linghaosong/Sextans}.
\end{abstract}


\maketitle

\thispagestyle{empty}

\section{Introduction}
\label{sec:intro}
Natural and scientific data are often
sparse and
large-scale which are stored as sparse
matrices.
The sparse matrices encode properties of nodes
the connection between nodes.
Sparse-matrix dense-matrix multiplication (SpMM) is a key computing routine
in a wide rage of 
applications, such as social networks~\cite{naumov2019deep}, chemical
reactivity prediction~\cite{coley2019graph}, drug design~\cite{cherkasov2014qsar},
security analysis~\cite{yamaguchi2014modeling}, 
sparse deep neural networks~\cite{han2015deep,wen2016learning,han2015learning},
and graph based machine learning \cite{hamilton2017inductive,kipf2016semi,xu2018powerful}.
SpMM performs the computation of 
$\mathbf{C} = \alpha\mathbf{A}\times\mathbf{B} + \beta\mathbf{C}$, where $\mathbf{A}$ is a sparse matrix,
$\mathbf{B}$ and $\mathbf{C}$ are dense
matrices, and $\alpha$ and $\beta$ are two
constant scalars. 
SpMM acceleration~\cite{yang2018design,gale2020sparse,huang2020ge,yan2020hygcn,geng2020awb,zhu2019sparse,han2016eie,wang2021gnnadvisor}
is attractive to
researchers of computer systems and architectures.

Application-specific accelerators 
boost the performance of  applications
in many domains.
However, the design of the
SpMM accelerator faces many challenges.

\noindent $\bullet$ {\bf Challenge 1 -- } Workload imbalance makes
SpMM difficult for parallelization. 
Row based parallelization~\cite{su2012clspmv,bell2009implementing,williams2007optimization} assigns
the processing of one row as a task for a processing engine or a thread (block). However,
because of the random distribution of non-zeros
in each row, 
processing engines 
with early completion time
will be idle.
To overcome the
workload imbalance, non-zero based parallelization~\cite{liu2015csr5,srivastava2020tensaurus,fowers2014high} 
(or similarly edge-enteric processing in graph processing acceleration~\cite{roy2013x,song2018graphr,zhu2015gridgraph,chi2016nxgraph}) is presented. However,  non-zero based parallelization
may incur
the read-after-write dependency
at the accelerator microarchitecture level, which leads to
a larger initial interval (II) for the scheduling.

\noindent $\bullet$ {\bf Challenge 2 -- } Inefficient memory accessing is another challenge.
Since the matrices of SpMM are large and can not be fit on chip, they are stored in off-chip memory. The processing of SpMM incurs random 
read accessing to
matrix $\mathbf{A}$, matrix $\mathbf{B}$, and matrix $\mathbf{C}$, and random 
write accessing to $\mathbf{C}$. It is dramatically inefficient to issue 
the huge amount of random accesses to off-chip memory.

\noindent $\bullet$ {\bf Challenge 3 -- } How to design a general-purpose accelerator which does not need to be rerun the time-consuming
flow of
synthesis/place/route. 
While many accelerators
have been designed for boosting computing performance and efficiency
in many application domains such as 
deep learning~\cite{wang2021autosa,cong2018polysa,zhang2015optimizing,ding2019req,zhang2017improving,arora2021tensor,song2017pipelayer,song2020accpar,song2019hypar,chen2014dadiannao,chen2016eyeriss,genc2021gemmini,shen2017maximizing,shen2017escher,shao2019simba},
dense linear algebra~\cite{de2020fblas,wang2021autosa,cong2018polysa,genc2021gemmini,de2020flexible},
graph processing~\cite{chi2021extending,ham2016graphicionado,ahn2015scalable,zhou2016high,dai2016fpgp,dai2018graphh,zhou2019hitgraph,song2018graphr,zhang2018graphp,zhuo2019graphq}, 
genomic and bio analysis~\cite{chen2020parc,huangfu2019medal,chen2020blink,chen2019lanmc,wu2019fpga,cali2020genasm,fujiki2018genax,turakhia2018darwin,guo2019hardware},
data sorting~\cite{samardzic2020bonsai,qiao2021fans,chen2019sorting,jun2017terabyte},
most are designed for {\it one specific
problem with fixed input and output size}. 
For FPGA accelerators even with improved tools such as
\cite{wang2021autosa,chi2021extending},
a new design will still consume
many hours or even a few days
due to long synthesis and 
place/route time. Moreover, it is a nightmare
for end users who are not an accelerator expert
to customize and rerun the flow 
to generate the accelerators for their applications.

In this paper we present \sextans
a streaming accelerator for accelerating
general-purpose
sparse-matrix den-matrix multiplication.
The contributions include:

\noindent $\bullet$ A hierarchical SpMM accelerator architecture. At the highest level,
  \sextans consists of (1) processing engine groups (PEGs), (2) modules to
  stream in/out matrices
  from/to off-chip HBM, and (3) modules to collect and perform element-wise multiplication and addition
  to obtain an updated $\mathbf{C}$. 
  PEGs and processing engines (PEs) are
  the key processing modules.
  A PEG consists of PEs and a PE consists of 
  processing units.
  
\noindent $\bullet$ PE-aware non-zero scheduling for a balanced workload with an II=1 pipeline.
The non-zero scheduling
of \sextans is an out-of-order~\cite{tomasulo1967efficient} scheduling.
The key idea of the scheduling is that a non-zero
is scheduled at the earliest cycle satisfying
that the row index of the scheduled non-zero
has no read-after-write (RAW) with the row index of non-zeros
being processed in previous $D$ (the distance of RAW dependency of a specific hardware
platform) cycles. 
The scheduling leads to an II=1 pipeline.
Similar to prior works~\cite{srivastava2020tensaurus,fowers2014high,zhu2015gridgraph}, we incorporate the scheduling in the preprocessing of the spares elements.

\noindent $\bullet$ Multi-level memory optimizations with HBM for efficient accessing and streaming. We partition the three matrices
of SpMM to fit the processing engines and on-chip memory (URAMs and BRAMs). For the off-chip memory,
matrices are streamed in/out in batches 
with a window size so the off-chip memory accessing
is always sequential. 
We partition
the random memory read and write
into a specific window,
so random memory accessing is
limited to on-chip fast memory.

\noindent $\bullet$ Hardware flexibility (HFlex) to support execution of different SpMMs 
directly by hardware
as a general-purpose accelerator.
 We enable the HFlex feature with
an iteration pointer list $\mathbf{Q}$ (similar to an instruction queue). 
We partition 
an arbitrary sparse matrix
$\mathbf{A}$ into multiple
$\mathbf{A}$ submatrices. 
We convert each $\mathbf{A}$ submatrix into a 
list of scheduled non-zeros. 
We store
the scheduled non-zero lists of 
all $\mathbf{A}$ submatrices 
linearly in a memory space.
We use an iteration pointer list $\mathbf{Q}$ to record the starting
index of each scheduled non-zero list. In the processing,
entries of $\mathbf{Q}$ serve as the loop iteration number and,
as a result, 
we support the execution of
an arbitrary SpMM without 
modification on the hardware.
 \sextansFPGA can be easily invoked
by OpenCL runtime
without handling the hardware and design details.

\noindent $\bullet$ Comprehensive evaluation.
We present
an FPGA prototype \sextansFPGA which is executable on a Xilinx U280 HBM FPGA board and a projected prototype \sextansASIC with higher bandwidth competitive to V100 and more frequency optimization.
We conduct comprehensive evaluation on 1,400 SpMMs on 200 sparse matrices. We compare Sextans with NVIDIA K80 and V100 GPUs.
\sextansFPGA achieves a 2.50x geomean speedup over K80 GPU and \sextansASIC achieves a 1.14x geomean speedup over V100 GPU (4.94x over K80).

\section{Background and Motivation}
\label{sec:background}
\subsection{Sparse-Matrix Dense-Matrix Multiplication}

SpMM performs the computation of 
$\mathbf{C} = \alpha\mathbf{A}\times\mathbf{B} + \beta\mathbf{C}$, where $\mathbf{A}$ is a sparse matrix,
$\mathbf{B}$ and $\mathbf{C}$ are dense
matrices, and $\alpha$ and $\beta$ are two
constant scalars. In algorithm modeling,
the sparse matrix $\mathbf{A}$ represents a graph,
the dense matrix $\mathbf{B}$ depicts 
the feature vectors of nodes,
and the dense matrix $\mathbf{C}$ is involved if
both old and new features are modeled.
Because natural
and social data structures are large and sparse
SpMM is a very useful computing
routine in many application domains.

For example, in sparse deep neural networks~\cite{han2015deep,wen2016learning,han2015learning},
matrix $\mathbf{A}$
represents
the pruned weight and
matrix $\mathbf{B}$
represent
feature maps, so the inference
is performed by $\mathbf{C} = 1.0\cdot\mathbf{A}\times\mathbf{B} + 0.0\cdot\mathbf{C}$. 
In graph based machine learning \cite{hamilton2017inductive,kipf2016semi,xu2018powerful}, 
matrix $\mathbf{B}$
represent
the node properties and
matrix $\mathbf{A}$ represents
the graph, so SpMM performs the graph
propagation.
Therefore,
researchers of computer systems and architectures
find
SpMM acceleration~\cite{yang2018design,gale2020sparse,huang2020ge,yan2020hygcn,geng2020awb,zhu2019sparse,han2016eie,wang2021gnnadvisor}
attractive.

\subsection{Sparse Matrix  Multiplication Acceleration}

\begin{figure}[tb]
\centering
\includegraphics[width=0.97\columnwidth]{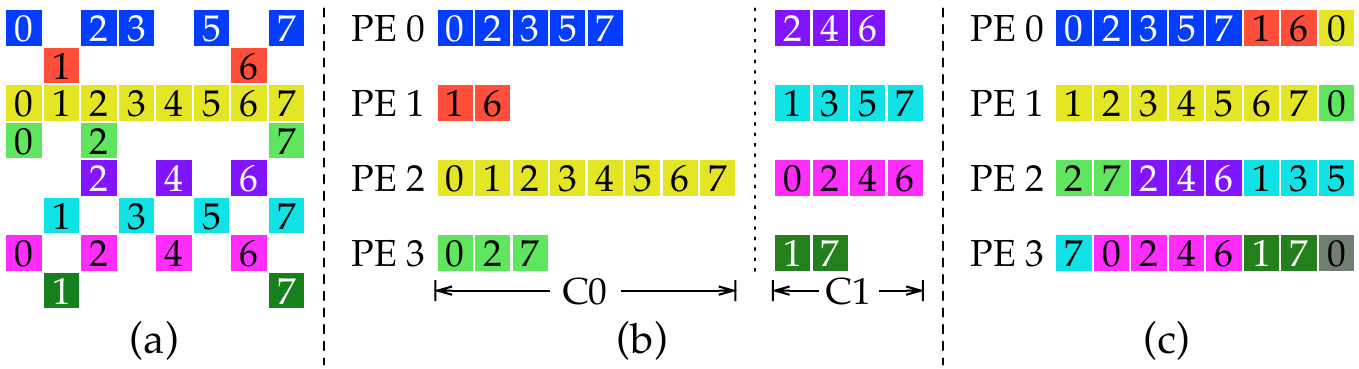}
\vspace{-9pt}
\caption{(a) a sparse matrix, (b) row based parallelization, and (c) non-zero based parallelization.}
\label{figure:sparse_parallel}
\vspace{-15pt}
\end{figure}

SpMM is challenging for parallelization.
Suppose we are using four processing engines (PEs) or threads to compute on 
a sparse matrix as shown in Figure~\ref{figure:sparse_parallel} (a) 
where the non-zeros (of matrix $\mathbf{A}$)
in row are colored the same.
The number on a square is the column index of
each non-zero.
An intuitive approach to paralleling SpMM on CPUs/GPUs
is row based parallelization~\cite{su2012clspmv,bell2009implementing,williams2007optimization} as shown in Figure~\ref{figure:sparse_parallel} (b).
It takes 2 steps C0 and C1 to process
the eight rows by the four PEs. However, there is
PE workload imbalance. For example, 
PE2 takes 8 cycles to process the yellow row
but other PEs finish their rows earlier and 
become idle in C0. The workload imbalance
wastes computing resources. To overcome the
workload imbalance, non-zero based parallelization~\cite{liu2015csr5,srivastava2020tensaurus,fowers2014high} 
(or similarly edge-enteric processing in graph processing acceleration~\cite{roy2013x,song2018graphr,zhu2015gridgraph})is presented.
As shown in 
Figure~\ref{figure:sparse_parallel} (c),
non-zeros of mutiple rows 
are packaged into a segment and 
the segments are assigned to PEs. 
The organization of non-zeros
is implemented as explicit or implicit
data formats~\cite{liu2015csr5,srivastava2020tensaurus,fowers2014high} which also encode the scheduling
of computing tasks.
Because the segments are 
equal in length, the PE workload is balanced. 
However, those techniques can not
be directly adopted in FPGA accelerators because pipeline and
memory related issues occur, and
details
are discussed in
Section~\ref{sec:mot}.

\subsection{High Bandwidth Memory}

High bandwidth memory (HBM)~\cite{hbmjedec} is designed for
applications which demands for high memory bandwidth.
HBM provides many pseudo and/or physical channels for 
channel-level parallel accessing and thus delivers
a high total bandwidth. For example, Xilinx U280 FPGA 
accelerator card~\cite{u280} is equipped with an HBM which
offers 32 pseudo channels. The bandwidth of each 
pseudo channel is 14.375 GB/s, for a total bandwidth of
460 GB/s. Because HBM is a new feature to
FPGAs, existing studies of FPGA HBM 
mainly focus on tool development~\cite{chi2021extending,guo2021autobridge,choi2021hbm} and benchmarking~\cite{huang2021shuhai,choi2020hls}, but very few applications. 
SpMM, a memory-intensive application
which is distinguished from
typical computation-intensive
FPGA applications~\cite{wang2021autosa,cong2018polysa,zhang2015optimizing,ding2019req,zhang2017improving,arora2021tensor}, is a good fit
for HBM. This work presents one of the first real application
for HBM FPGA.

\subsection{Motivation}
\label{sec:mot}

Our work aims at achieving the following
goals to addressing limitations in prior works.

\noindent
{\bf $\bullet$ A general-purpose and user-friendly SpMM accelerator.} Domain specific architectures~\cite{cong2014accelerator,cong2019customizable,hennessy2019new,dally2020domain}
have been designed for boosting computing performance and efficiency
in many application domains such as 
deep learning~\cite{wang2021autosa,cong2018polysa,zhang2015optimizing,ding2019req,zhang2017improving,arora2021tensor,song2020accpar,chen2014dadiannao,chen2016eyeriss,genc2021gemmini,shen2017maximizing,shen2017escher,shao2019simba,song2017pipelayer,song2019hypar,hsu2021accelerating},
dense linear algebra~\cite{de2020fblas,wang2021autosa,cong2018polysa,genc2021gemmini,de2020flexible},
graph processing~\cite{chi2021extending,ham2016graphicionado,ahn2015scalable,zhou2016high,dai2016fpgp,dai2018graphh,zhou2019hitgraph,song2018graphr,zhang2018graphp,zhuo2019graphq,hu2021graphlily,besta2021sisa,matam2019graphssd}, 
genomic and bio analysis~\cite{chen2020parc,huangfu2019medal,chen2020blink,chen2019lanmc,wu2019fpga,cali2020genasm,fujiki2018genax,turakhia2018darwin,guo2019hardware,cali2020genasm}, and
data sorting~\cite{samardzic2020bonsai,qiao2021fans,chen2019sorting,jun2017terabyte}.
However, most accelerators are designed for one specific
problem with fixed input and output size. 
To support a different problem 
configuration, the accelerator 
architecture is to be modified, which
will consumes several weeks to months
for architecture re-design and
chip tape-out. For FPGA accelerators,
even with improved tools such as
\cite{wang2021autosa,chi2021extending},
a new design will still consume
many hours even a few days
due to long synthesis and 
place/route time. 
Another method is to
design a kernel for a
fixed-size problem and
decompose a new problem to
multiple kernels of the
fixed-size problem and launch those kernels by the runtime.
However,
the runtime 
overhead is significant.
For example, we can build
a accelerator for a matrix
multiplication with a fixed size of $4096\times 4096$ with~\cite{wang2021autosa} 
and we map
50 SNAP matrices~\cite{snapnets}
where the row/column number ranges from
1,005 to 456,626 and the number of non-zeros
ranges from 20,296 to 14,855,842.
The average number of decomposed fixed-size kernels
for the SNAP matrices is 1793.
The OpenCL runtime overhead for
launching one kernel is around 0.15ms.
So the average runtime overhead
for launching kernels is 269ms.
As a comparison, the average execution time
of SpMM on the SNAP matrices by an NVIDIA K80
GPU is 5.85ms.
Moreover, it is a nightmare
for end users who are not accelerator experts
to customize and rerun the flow for their applications.
As we discussed before, SpMM is an abstraction
kernel for many applications. 
\sextans features
the hardware flexibility (HFlex) to 
directly
support execution of different SpMMs 
by hardware as
a general-purpose and democratized SpMM accelerator.
Users can deploy
\sextansFPGA for different 
problems without
rerunning the synthesis/place/route flow and 
they can easily invoke \sextansFPGA 
in their applications by OpenCL runtime
without handling the hardware details.

\noindent
{\bf $\bullet$ PE-aware non-zero scheduling for a
balanced workload.} Although 
non-zero based parallelization~\cite{liu2015csr5,fowers2014high,roy2013x,zhu2015gridgraph} alleviates the load
imbalance issue, directly applying non-zero 
based parallelization to accelerators incurs
a dependency issue 
on the microarchitecture.
For example, in Figure~\ref{figure:sparse_parallel} (c)
the blue elements held by PE 0
are accumulated to the same destination 
and the floating-point
ADD usually takes 7 to 10 cycles (depending on 
specific FPGAs). Thus, a
read-after-write (RAW) conflict 
occurs for blue elements 2, 3, 5, 7 and HLS tools will
schedule a large initiation interval (II), leading to 
long computing latency. To overcome this RAW 
dependency issue, we present a PE-aware no-zero 
scheduling. The key idea is to apply out-of-order~\cite{tomasulo1967efficient} scheduling
to move back a conflict element and move forward a non-conflict element within a scheduling window
to resolve the RAW dependency.

\noindent
{\bf $\bullet$ Multi-level memory optimizations for efficient accessing and  streaming.}
SpMM face three memory related challenges: 
({\bf C1}) bank conflict, 
({\bf C2}) irregular memory accessing, 
and ({\bf C3}) off-chip large matrix accessing.

{\bf C1} -- A bank conflict happens when two or more
processing units access the same bank.
For example, 
in Figure~\ref{figure:sparse_parallel} (c)
PE 0 and PE 1 both need to read the element 
with column index 0 at Cycle 8.
A bank
conflict occurs if the the memory storing element 0
 has only one port. 
To overcome the bank conflict, we duplicate
the read-only matrix shard to the PEs.

{\bf C2} -- The irregular column index shown as colored square numbers
in Figure~\ref{figure:sparse_parallel}
(b) and (c) lead to irregular memory read 
requests, whereas the
irregular row destination of PEs
in Figure~\ref{figure:sparse_parallel} (c)
leads to irregular memory write 
requests. Although our accelerators are equipped 
with HBM which has higher memory bandwidth, 
the latency of accessing HBM is still high
(up to 100 cycles)~\cite{choi2021hbm}.
Inspired by the idea of caching random accessing on
a higher memory hierarchy in graph processing ~\cite{song2018graphr,zhu2015gridgraph},
we partition the random memory read and write into a specific window, so random memory accessing is limited to on-chip fast memory.
We store read-only dense
in BRAMs to limit random read on chip.
We use a scratchpad memory (FPGA URAMs)
to limit random accumulation
(read and write) on chip.
We also achieve an II=1 scheduling 
that will further
hide on-chip accessing latency.

{\bf C3} -- The three matrices $\mathbf{A}$,
$\mathbf{B}$, and $\mathbf{C}$ in SpMMs
are so large that we can not store them
on chip. For example, one evaluated
matrix {\tt bundle\_adj} consumes 3.2 GB
memory footprint but the total on-chip memory (SRAM)
of a Xilinx U280 FPGA is 41 MB.
We partition the three matrices
according to the processing window size and
store the partitioned matrix shards in HBM.
We do not issue individual element accessing
to HBM but only read or write a matrix shard.
This allows HBM to be streamed for efficient accessing.


\section{\sextans Architecture}
\label{sec:architecture}
\subsection{Overall Architecture}

\subsubsection{Overall Architecture}

\begin{figure}[tb]
\centering
\includegraphics[width=0.8\columnwidth]{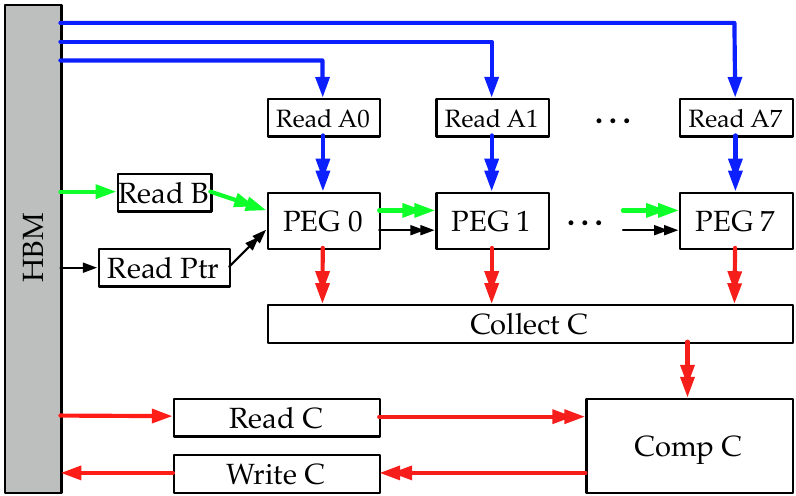}
\vspace{-9pt}
\caption{\sextans overall architecture.}
\label{figure:overall}
\vspace{-15pt}
\end{figure}

Figure~\ref{figure:overall}
illustrates
the overall architecture 
of \sextans.
An arrow demotes the data transfer direction and a double-arrow indicates
a FIFO connection. Data transfers of
$\mathbf{A}$, $\mathbf{B}$, and $\mathbf{C}$
are colored in 
blue, green, and red respectively.
We deploy
8 processing engine groups (PEG 0 -- 7) 
to compute $\mathbf{C}_{AB} = \mathbf{A}\times\mathbf{B}$. Each PEG contains
8 PEs. 
To stream in and supply
the disjoint 
partitioned matrix $\mathbf{A}_{pj}$ 
(defined in Equation~\ref{eq:AB_p})
from HBM
to each PEG,
we disseminate
8 Read A modules.
We deploy
a Read B module to
stream in a window of matrix
$\mathbf{B}_{ji}$ from HBM
and broadcast
$\mathbf{B}_{ji}$ to the 8 PEGs.
Each PEG also serves as a relaying node
to form a chain-based broadcasting network.
We do not use a one-to-all broadcasting network
because a one-to-all broadcasting leads to
low frequency~\cite{cong2018latte} and route failure.
We deploy
a Read Ptr module 
to deliver pointers $\mathbf{Q}$ of
out-of-order scheduled
non-zeros
to PEGs.
A chain-based broadcasting network
delivers the pointers to PEGs.
A Collect C module collects
the disjoint $\mathbf{C}_{\alpha AB}|_p$. 
A Comp C module
performs the element-wise computation of
$\mathbf{C}_{out} = \mathbf{C}_{\alpha AB} + \beta\cdot\mathbf{C}_{in}$ where
$\mathbf{C}_{AB}$ is supplied by the Collect C model,
$\mathbf{C}_{in}$ is supplied by a Read C module which
streams in $\mathbf{C}_{in}$ from HBM, and
$\mathbf{C}_{out}$ is sent to a Write C module
to be streamed out to HBM.
We assign
1 HBM channel to 
pointers $\mathbf{Q}$, 
4 HBM channels to 
matrix $\mathbf{B}$,
8 HBM channels to 
matrix $\mathbf{A}$,
8 HBM channels to 
matrix $\mathbf{C}_{in}$, and
8 HBM channels to 
matrix $\mathbf{C}_{out}$.

\subsubsection{Overall Processing Logic}

\begin{figure}[tb]
\centering
\includegraphics[width=0.85\columnwidth]{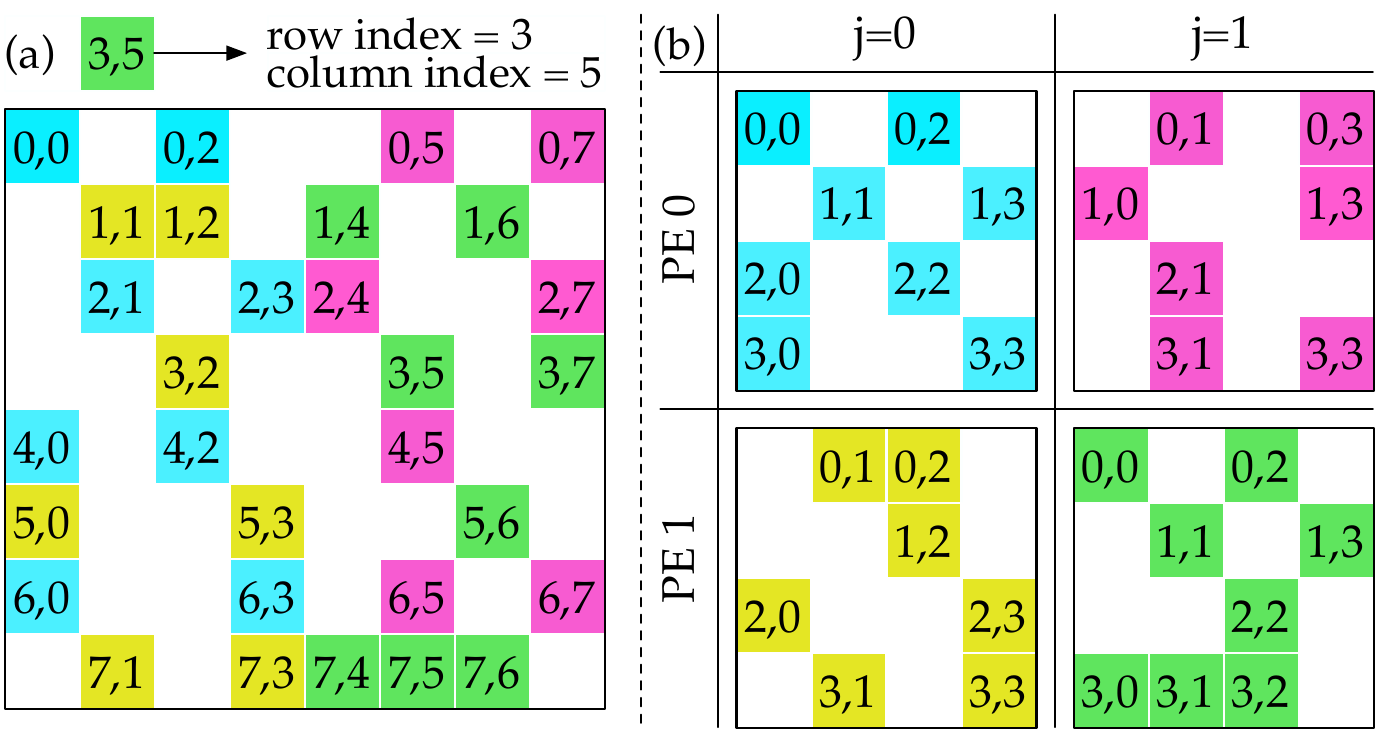}
\vspace{-9pt}
\caption{(a) A sparse matrix example and (b) its partitioning.}
\label{figure:partition_example}
\vspace{-15pt}
\end{figure}

We discuss the overall logic of how we partition the spares and dense
matrices and how we process
SpMM.
We separate the computation of
$\mathbf{C} = \alpha\mathbf{A}\times\mathbf{B} + \beta\mathbf{C}$ into three phases,
\vspace{-3pt}
\begin{equation} 
\label{eq:ABC}
\left\{
\begin{array}{l}
\mathbf{C}_{AB} = \mathbf{A}\times\mathbf{B} \\
\mathbf{C}_{\alpha AB} = \alpha\cdot\mathbf{C}_{AB} \\
\mathbf{C}_{out} = \mathbf{C}_{\alpha AB} + \beta\cdot\mathbf{C}_{in}
\end{array}.
\right.
\vspace{-0pt}
\end{equation} 
$\mathbf{C}_{in}$, $\mathbf{C}_{AB}$($\mathbf{C}_{\alpha AB}$) and $\mathbf{C}_{out}$ are
the input $\mathbf{C}$ matrix, 
the intermediate multiplication $\mathbf{C}$ matrix,
and the output $\mathbf{C}$ matrix
respectively. The computation of 
$\mathbf{C}_{AB} = \mathbf{A}\times\mathbf{B}$
is the most challenging phase in the SpMM acceleration.
With the intermediate multiplication 
$\mathbf{C}_{AB}$ we perform the element-wise
multiplication with $\alpha$ to obtain
$\mathbf{C}_{\alpha AB}$.
Then
we stream in $\mathbf{C}_{in}$
from off-chip memory,
execute element-wise multiplication/addition
for $\mathbf{C}_{out} = \mathbf{C}_{\alpha AB} + \beta\cdot\mathbf{C}_{in}$, and stream out 
$\mathbf{C}_{out}$ to off-chip memory.

$\mathbf{A}$, $\mathbf{B}$, and $\mathbf{C}$ are large matrices
that do not fit on chip, thus we need to partition
the three matrices and reform the computation
of $\mathbf{C}_{AB} = \mathbf{A}\times\mathbf{B}$.
The dimension of 
$\mathbf{A}$, $\mathbf{B}$, and $\mathbf{C}$
are $M\times K$, $K\times  N$ and $M\times N$ respectively. 
First, we partition $\mathbf{B}$ rows. 
We partition each row into segments
with a length of $N_0$.
So $\mathbf{B}$ becomes $N/N_0$
submatrices $\mathbf{B}_i$ with
the dimension of $\mathbf{B}_i$ set to $K\times N_0$. 
The multiplication of $\mathbf{C}_{AB} = \mathbf{A}\times\mathbf{B}$
changes to:
\vspace{-3pt}
\begin{equation} 
\label{eq:AB_new}
\left\{
\begin{array}{l}
\mathbf{C}_{AB} =
\{\mathbf{C}_{AB_i}|i\in\{0,1,...,N/N_0-1\}\} \\
\mathbf{C}_{AB_i} =  \mathbf{A}\times\mathbf{B}_i
\end{array}.
\right.
\vspace{-3pt}
\end{equation} 
Next we partition $\mathbf{B}_i$ columns.
Each column converts to
$K/K_0$ column segments with a length of $K_0$.
Because the $\mathbf{A}$ row is associated
the $\mathbf{B}_i$ column, 
we divide the $\mathbf{A}$ row
into row segments.
These segments
have a length of $K_0$
(also referred to as window size in the following content).
Thus, $\mathbf{B}_i$ is partitioned into
$K/K_0$ submatrices $\mathbf{B}_{ji}$ and
$\mathbf{A}$ is partitioned into
$K/K_0$ submatrices $\mathbf{A}_{j}$.
The computation of 
$\mathbf{C}_{AB_i} =  \mathbf{A}\times\mathbf{B}_i$
becomes:
\vspace{-3pt}
\begin{equation} 
\label{eq:AB_i}
\left\{
\begin{array}{l}
\mathbf{C}_{AB_i} = 
\sum_j^{K/K_0}\mathbf{C}_{A_jB_{ji}}\\
\mathbf{C}_{A_jB_{ji}} = \mathbf{A}_j\times\mathbf{B}_{ji}\\
\end{array}.
\right.
\vspace{-3pt}
\end{equation} 
$\mathbf{A}_j$ is a sparse matrix and the
uneven distribution of non-zeros leads
to the workload imbalance issues as 
discussed earlier. 
In order to achieve an
approximately uniform probability distribution of the non-zeros
we split
non-zeros of $\mathbf{A}_j$ into
$P$ bins. The bin $p$ holds
non-zeros whose row index $row$ satisfies
$(row~\text{mod}~P) == p$.
So $\mathbf{A}_j$ transforms into
$P$ submatrices  $\mathbf{A}_{pj}$ and
we change the computation of 
$\mathbf{C}_{A_jB_{ji}} = \mathbf{A}_j\times\mathbf{B}_{ji}$ to:
\vspace{-3pt}
\begin{equation} 
\label{eq:AB_p}
\left\{
\begin{array}{l}
\mathbf{C}_{A_jB_{ji}} =  
\{\mathbf{C}_{A_{pj}B_{ji}}|p\in\{0,1,...,P-1\}\}
\\
\mathbf{C}_{A_{pj}B_{ji}} = \mathbf{A}_{pj}\times\mathbf{B}_{ji}\\
\end{array}.
\right.
\vspace{-3pt}
\end{equation} 

The partitioning of the three matrices
determines the
coarse-grained scheduling for
$\mathbf{C}_{AB} = \mathbf{A}\times\mathbf{B}$. 
Equation~\ref{eq:AB_new}
and Equation~\ref{eq:AB_i} are processed
sequentially and
Equation~\ref{eq:AB_p} is performed
in parallel.
In \sextans architecture, we pass  
parameters $N$ and $K$ to the accelerator.
Then we calculate
the iteration number $N/N_0$
of Equation~\ref{eq:AB_new} and
the iteration number $K/K_0$ 
of Equation~\ref{eq:AB_i}.
Two outer loops schedules
Equation~\ref{eq:AB_new} and
Equation~\ref{eq:AB_i}.
We set up a total number of $P$ parallel
processing engines (PEs) to perform
Equation~\ref{eq:AB_p}.
We assign the submatrix multiplication $\mathbf{C}_{A_{pj}B_{ji}} = \mathbf{A}_{pj}\times\mathbf{B}_{ji}$
to the $p$-th PE.
Because the iterator $p$ (introduced by
Equation~\ref{eq:AB_p}) is only associated
with the $p$-th PE, 
$\mathbf{A}_{pj}$ and $\mathbf{C}_{A_{pj}B_{ji}}$ of each PE
are disjoint.
However, 
the iterator $i$ (introduced by
Equation~\ref{eq:AB_new}) and $j$
(introduced by Equation~\ref{eq:AB_i})
are exposed to every PE, so 
every PE accesses $\mathbf{B}_{ji}$.

We show an example of sparse matrix
and its partitioning in
Figure~\ref{figure:partition_example}, assuming we have 2 PEs, 
and window size is 4.
In iteration $j=0$, we assign
the elements
(colored in blue and yellow)
with column index 0 - 3 
to the two PEs. 
In iteration $j=1$, we assign
the elements
(colored in magenta and green)
with column index 4 - 7 
to the two PEs. 
For PE 0,
we designate the
elements with row index 0, 2, 4, 6
of 
Figure~\ref{figure:partition_example} (a).
For PE 1,
we designate the
elements with row index 1, 3, 5, 7
of 
Figure~\ref{figure:partition_example} (a).
The original row index is scheduled
to be interleaved (mod $P$) to achieve 
statistically
even distribution of non-zeros.
Note that both row index
and column index are compressed.
For example, we convert
the green element
(3,5) to green
(1,1) in iteration $j=1$ for PE 1.

\subsection{Processing Engine}

\begin{figure}[tb]
\centering
\includegraphics[width=0.975\columnwidth]{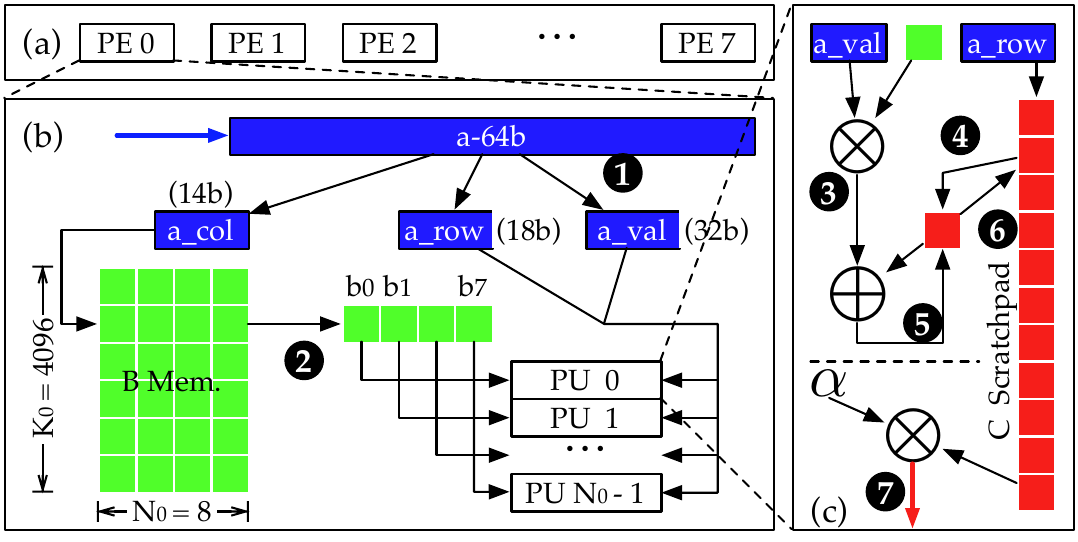}
\vspace{-9pt}
\caption{The architecture of (a) a PEG, (b) a PE, and (c) a PU.}
\label{figure:peg}
\vspace{-15pt}
\end{figure}

\begin{figure*}[tb]
\centering
\includegraphics[width=2.05\columnwidth]{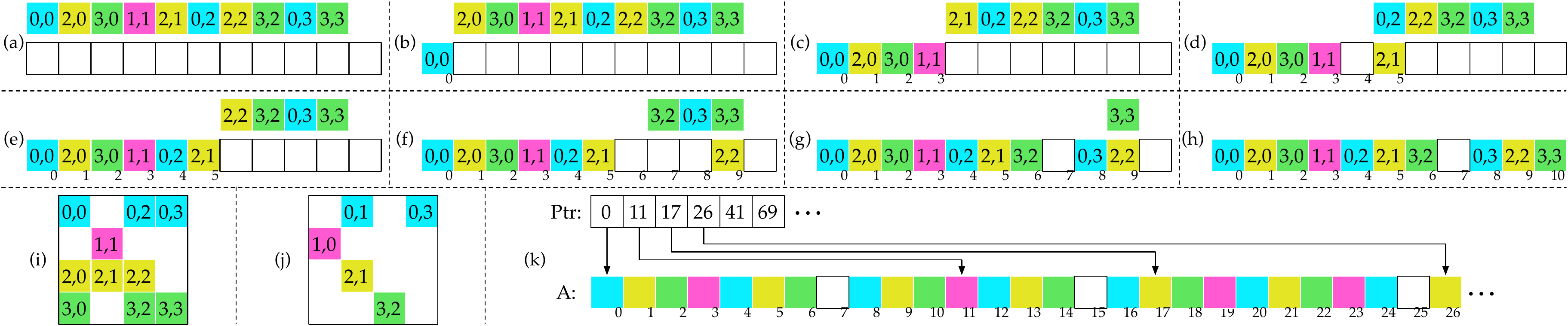}
\vspace{-3pt}
\caption{(a -- h) The non-zero scheduling for the example (i). (i, j) The non-zero graphs for
two example
submatrix $\mathbf{A}_{pj}$. (k)~Supporting HFlex 
with a pointer array which 
records the starting index
of scheduled non-zeros of
$\mathbf{A}_{pj}$ submatrices.}
\label{figure:scheduling}
\vspace{-9pt}
\end{figure*}

Figure~\ref{figure:peg} shows the architecture of a PEG, a PE,
and a processing unit (PU). 
A PEG contains
8 PEs and a PE is the key module in
\sextans architecture. A PE contains 
$N_0 = 8$ PUs. A PU performs the
computation related to one non-zero 
scalar of matrix $\mathbf{A}_{pj}$.

To better illustrate the architecture of
PE and PU, 
we change the
submatrix multiplication $\mathbf{C}_{A_{pj}B_{ji}} = \mathbf{A}_{pj}\times\mathbf{B}_{ji}$
of $p$-th processing engine into
sparse scalar format to:
\vspace{-3pt}
\begin{equation} 
\label{eq:pe}
\left\{
\begin{array}{l}
c_{kq} ~+= a_{kl}\cdot b_{lq}\\
\forall a_{kl}\in\vec{u}_l,~a_{kl}\neq0,~\forall \vec{u}_l\in\mathbf{A}_{pj},~\vec{u}_l\neq\vec{0}\\
c_{kq}\in\mathbf{C}_{A_{pj}B_{ji}},~b_{lq}\in\mathbf{B}_{ji}\\
\end{array},
\right.
\vspace{-3pt}
\end{equation} 
where $c_{kq}$, $a_{kl}$, and $b_{lq}$ is a scalar(non-zero)
of $\mathbf{C}_{A_{pj}B_{ji}}$, $\mathbf{A}_{pj}$, and $\mathbf{B}_{ji}$ respectively. $\vec{u}_l$ is a column
vector of $\mathbf{A}_{pj}$.
We iterate on 
the non-zeros $a_{kl}$ of the column vector. 
Thus the processing is an outer-product like manner~\cite{pal2018outerspace}. However, a
RAW dependency conflict may happen when a
non-zero
$a_{kl}$ from the 
next column vector has the same
row index as a non-zero which is 
being processed.
Thus, a RAW dependency conflict happens
and the HLS can not achieve
an II=1 scheduling.
We will present
a PE-aware scheduling to resolve this issue in Section~\ref{sec:nonzero-schedule}.
The iterator $q$ on the second dimension
of $\mathbf{C}_{A_{pj}B_{ji}}$ does not rely on
any dependency, so 
we can schedule
the processing on $q$ 
in parallel. The length of the 
second dimension
of $\mathbf{C}_{A_{pj}B_{ji}}$ is $N_0=8$.

A PE consumes one non-zero of $\mathbf{A}_{pj}$
and performs the multiplication and accumulation
for $N_0$ parallel elements of
$\mathbf{B}_{ji}$ and 
$\mathbf{C}_{A_{pj}B_{ji}}$ in an II=1 pipeline as shown in Figure~\ref{figure:peg} (b).
One non-zero originally
consumes 96 bits where
each of row index, column index,
and floating-point 
value consumes 32b its.
Because the sparse matrix is
partitioned, we can compress
the row and column index to save
memory footprint.
We encode the row index, column index, and value
of the non-zero in a 64-bit element {\tt a-64b}.
The first step \ballnumber{1} is to decode
{\tt a-64b} to a 14-bit column index {\tt a\_col},
a 18-bit row index {\tt a\_row}, and
a 32-bit floating-point value {\tt a\_val}.
{\tt a\_col} is indexed to on-chip B memory
and {\tt a\_row} is indexed to on-chip
C scratchpad memory.
The window size (depth) of B memory
is $K_0=4096$ which consumes 12-bit.
The depth of C scratchpad memory
is 12,288\footnote{We use a Xilinx U280 FPGA. The URAM size is 4096$\times$72 bit. 
With $N_0=8$, we allocate $12288/4096\times8/2=12$ URAMs for each PE. 768 (80\%) URAMs are consumed.} which consumes 14-bit. So 14/18 bits are sufficient for
{\tt a\_col}/{\tt a\_row}.
In the next step
\ballnumber{2}, {\tt a\_col} is used to
retrieve $N_0=8$ b elements 
$b_0$ to $b_7$ from the B memory.
$b_i$ is sent to the $i$-th PU.
The 8 PUs share
{\tt a\_row} and {\tt a\_val} from
step \ballnumber{1}.
Inside a PU (Figure~\ref{figure:peg} (c)), {\tt a\_val} is multiplied with
a b element $b_i$ \ballnumber{3}.
{\tt a\_row} is the index of the accumulation
C element $c_{kq}$ and 
is sent
to C scratchpad memory to fetch $c_{kq}$
\ballnumber{4}. \ballnumber{5} performs 
the accumulation of $c_{kq} ~+= a_{kl}\cdot b_{lq}$. \ballnumber{6} stores the updated
$c_{kq}$ back to C scratchpad memory.
After the computation of $\mathbf{C}_{A_{pj}B_{ji}} = \mathbf{A}_{pj}\times\mathbf{B}_{ji}$
 for one window is done, we iterate 
 and process the next window (Equation~\ref{eq:AB_i}).
After the computation of Equation~\ref{eq:AB_i}
is done, we perform the element-wise computation of
$\mathbf{C}_{\alpha A_{pj}B_{ji}} = \alpha\cdot\mathbf{C}_{A_{pj}B_{ji}}$. In the step \ballnumber{7},
the C elements are streamed out from
the C scratchpad memory and multiplied
with a constant $\alpha$ element
by element.
The multiplied
results $\mathbf{C}_{\alpha A_{pj}B_{ji}}$ are sent to the Collect C module.
Note that the HLS schedules
some steps of \ballnumber{1} -
\ballnumber{7} to be processed concurrently.

\subsection{Non-zero Scheduling}
\label{sec:nonzero-schedule}

The non-zero scheduling is platform-specific and PE-aware 
because the scheduling is based on
the distance $D$ of RAW dependency of a specific hardware
platform and the processing status of a PE.
Instead of in-order scheduling which is
unable to
fully utilize the pipeline, the non-zero scheduling
of \sextans uses an out-of-order~\cite{tomasulo1967efficient} scheduling.
The key idea of the scheduling is that
we schedule 
a non-zero
at the earliest cycle so
that the row index of the scheduled non-zero
has no RAW with the row index of non-zeros
being processed in previous $D$ cycles. 
The scheduling leads to an II=1 pipeline.
Similar to prior works~\cite{srivastava2020tensaurus,fowers2014high,zhu2015gridgraph}, we incorporate the scheduling in the preprocessing of the spares elements.
We provide 
the non-zero scheduling 
as a host C++ wrapper
for users.

We show an example of non-zero scheduling in
Figure~\ref{figure:scheduling} (a) - (h) for a
sparse matrix Figure~\ref{figure:scheduling} (i).
We assume the RAW dependency distance $D$ is 4 in this example.
The non-zeros which has the same row index 
are colored the same.
The non-zeros of Figure~\ref{figure:scheduling} (i)
is listed in column-major order in Figure~\ref{figure:scheduling} (a). 
In Figure~\ref{figure:scheduling} (b),
the first
non-zero, blue (0,0) -- (row index, column index)
is scheduled to Cycle 0.
For the next three non-zeros, 
we can safely
schedule them
to a following cycle because there is no
RAW conflict in Figure~\ref{figure:scheduling} (c).
In Figure~\ref{figure:scheduling} (d), yellow (2,1) 
conflicts with yellow (2,0) at Cycle 1 because $4-1<D=4$,
and it is scheduled to the earliest Cycle 5.
The blank(bubble) Cycle 4 is filled
by blue (0,2)
in Figure~\ref{figure:scheduling} (e).
Next, yellow (2,2) is scheduled to Cycle $5+4=9$
in Figure~\ref{figure:scheduling} (f).
In Figure~\ref{figure:scheduling} (g)
green (3,2) and blue (0,3)
is scheduled to Cycle 6 and Cycle 8 respectively.
The final non-zero green (3,3) is scheduled to
Cycle 10. Although the scheduling may
contain bubbles such as Cycle 7 in
Figure~\ref{figure:scheduling} (h),
bubbles are aggressively eliminated.
As a comparison, column-major in-order
scheduling consumes 15 cycles and
row-major in-order
scheduling consumes 28
for the example of Figure~\ref{figure:scheduling} (i).

\subsection{Hardware Flexibility
for Arbitrary SpMMs}
\begin{algorithm}[htb] 
\small
\caption{\sextans HFlex SpMM.} 
\label{alg:spmm} 
\begin{algorithmic}[0]
\REQUIRE ~~\\ 
(1) matrix $\mathbf{A}$, $\mathbf{B}$, and $\mathbf{C}$, (2) pointer list $\mathbf{Q}$, 
(3) constant $\alpha$ and $\beta$, and
(4) matrix hyperparameter $M$, $K$, and $N$.

\ENSURE ~~\\ 
$\mathbf{C} = \alpha\mathbf{A}\times\mathbf{B} + \beta\mathbf{C}$.\\
\end{algorithmic}
\vspace{-12pt}
\begin{algorithmic}[1]

\FOR{($0\leq i<N/N_0$)} 
\STATE $\mathbf{C}_{AB_i}  \leftarrow \mathbf{0}$
\FOR{($0\leq j<K/K_0$)} 
\STATE {\tt Read} $\mathbf{B}_{ji}$
\FORALLP{($0\leq p< P$)} 
\FOR{($\mathbf{Q}_j\leq r<\mathbf{Q}_{j+1}$)} 
\FORALLP{($0\leq q< N_0$)} 
\STATE $c_{kq} \leftarrow c_{kq} + a_{kl}\cdot b_{lq}$
\ENDFAP
\ENDFOR
\ENDFAP
\ENDFOR
\STATE $\mathbf{C}_i \leftarrow \alpha\mathbf{A}\times\mathbf{B}_i + \beta\mathbf{C}_i$
\ENDFOR
\end{algorithmic}
\end{algorithm}

\sextans features
the hardware flexibility (HFlex) to 
directly
support execution of different SpMMs 
by hardware as
a general-purpose SpMM accelerator.
We enable the HFlex feature with
a pointer list $\mathbf{Q}$ (similar to an instruction queue). An arbitrary sparse matrix
$\mathbf{A}$ is partitioned into multiple
submatrices $\mathbf{A}_{pj}$. Each $\mathbf{A}_{pj}$ is converted into a 
list of scheduled non-zeros. The scheduled non-zero lists of 
all submatrices $\mathbf{A}_{pj}$ are stored linearly in a memory space.
We use a pointer list $\mathbf{Q}$ to record the starting
index of each scheduled non-zero list. For example, we place the 
scheduled non-zero list of Figure~\ref{figure:scheduling} (i)
in the space of 0 - 10 as show in 
Figure~\ref{figure:scheduling} (l). The 
scheduled non-zero list of a following $\mathbf{A}$ submatrix
displayed in Figure~\ref{figure:scheduling} (k)
is placed in the space of 11 - 16. So we set $\mathbf{Q}_1=11$
and $\mathbf{Q}_2=17$. The first entry of $\mathbf{Q}$ is always set to 0. The number of entries in $\mathbf{Q}$ is $K/K_0 + 1$, because
$K_0$ is the length for partitioning
the whole sparse matrix $\mathbf{A}$ and 
$K_0$ is also the
window size for the out-of-order non-zero scheduling.

We present \sextans HFlex SpMM processing 
in Algorithm~\ref{alg:spmm}. The inputs to 
the algorithm are (1) matrix $\mathbf{A}$ (with non-zeros scheduled), 
matrix $\mathbf{B}$, matrix $\mathbf{C}$,
(3) pointer array $\mathbf{Q}$,
(3) two constants $\alpha$, $\beta$,
and (4) three hyperparameters $M$, $K$, $N$
which describe the shape/dimension of the SpMM.
Line 5 - 11 performs the core computation
of SpMM. 
We deploy 8 PEGs where each PEG contains 8 PEs. 
So the parallel factor for 
Line 5 is $P=64$. Line 6 - 10 is a PE region
where a PE uses the pointer list $\mathbf{Q}$
to determine the loop number for a specific
sparse submatrix $\mathbf{A}_{pj}$.
Inside a PE, we unroll the scalar 
computation for a factor of $N_0=8$
to share one sparse $\mathbf{A}$ scalar
with 8 dense $\mathbf{B}$ scalars. 
With \sextans HFlex SpMM processing 
method, the parameters passed to the 
hardware accelerator are fixed.
Specifically, memory pointers of
$\mathbf{A}$, $\mathbf{B}$, $\mathbf{C}$,
and $\mathbf{Q}$, and constant scalars
$M$, $K$, $N$, $\alpha$ and $\beta$
are passed to the accelerator. For
a different SpMM, the data of matrix $\mathbf{A}$, $\mathbf{B}$ and $\mathbf{C}$
only affects the contents stored in the memory
space specified by the memory pointers.
We pass the memory pointers and constant
scalars according to a specific SpMM
to the accelerator without changing the 
accelerator. Thus, \sextans supports
the HFlex feature.

\subsection{Discussion on Other Architectural Issues}

We discuss five architectural issues in this section.

\noindent $\bullet$ (1) {\bf Streaming in B matrix.} SpMM actually
issues random read to $\mathbf{B}$ which is stored
in off-chip HBM but we need
to alleviate the random read to off-chip memory. 
Matrix $\mathbf{B}$ is partitioned into windows
(with a window size $K_0$). In one logical cycle,
the random accessing only happens on a $\mathbf{B}$
window. So we stream in a $\mathbf{B}$ window 
(Line 4 of Algorithm~\ref{alg:spmm})
before invoking the 8 PEGs. Thus, the read accessing
to HBM is sequentially batched.

\noindent $\bullet$ (2) {\bf Initialization of C matrix.} Similar to the
situation of accessing $\textbf{B}$, we can not afford
the cost for random accessing $\mathbf{C}$
in off-chip memory. We keep an on-chip scratchpad
memory to accumulate $\mathbf{C}$. As a result, 
$\mathbf{C}$ must be initiated to be 0
(Line 2 of Algorithm~\ref{alg:spmm}).
Each \sextans PE performs
initiates
a disjoint set of $\mathbf{C}$
in parallel.

\noindent $\bullet$ (3) {\bf Irregular accessing on chip.}
One input a-64b~\ref{figure:peg} will issue
one random read 
access (indexed by a\_col) to $\textbf{B}$
and one random read and write access
(indexed by a\_row) to $\textbf{C}$.
The two random accesses happen on on-chip memory.
Although the latency for a specific access or
computation is larger than 1 and the latency
for processing one $\mathbf{A}$ element
is 15 cycles
on a Xilinx U280
FPGA, with \sextans non-zero scheduling we achieve an
II=1 pipeline.

\noindent $\bullet$ (4) {\bf Synchronization.} We do not place explicit
synchronization barriers for the PEs. Instead,
we use FIFO to form a loose 
synchronization which is implicitly implemented 
in the broadcasting where $\textbf{B}$ elements
are sent from one Read B module (producer) to PEs. 
The FIFO 
depth is 8. So at most 8 ahead/delay cycles of
asynchronization are 
tolerated between PEs. If the asynchronization cycles
are larger than 8, there is one PE (PEx) 
which has consumed
all elements of its FIFO and another PE (PEy) 
whose FIFO is full. At the producer side, the sending is stalled.
PEx keeps idle because the connected FIFO is empty.
After PEy consumes at least one element in 
the connected FIFO, the processing resumes.

\noindent $\bullet$ (5) {\bf Speedup breakdown.} 
To understand the speedup breakdown, in Table~\ref{table:speedup}
we use Matrix {\tt crystm03} as an example to show the incremental and accumulative speedups
with the increase of the optimizations applied.
For the baseline, we cache
dense matrix blocks, scream in sparse matrix in row order (CSR), and there is no sharing.
The 8 PUs relate to computation (sharing) optimization, the 64 PEs
relate to memory optimization, and
the OoO scheduling relates to both
computation and memory optimizations.

\begin{table}[tb]
\caption{Incremental and accumulative speedups with the increase of optimizations applied on {\tt crystm03}.}
  \label{table:speedup}
  \vspace{-9pt}
  \centering
  \small
  \begin{tabular}{ccccc}
    \hline
    & Baseline & OoO Scheduling & 8 PUs & 64 PEs \\
    \hline
    {\bf Incr.}   & $1\times$ & $9.97\times$   & $7.97\times$ & $45.3\times$ \\
    {\bf Accum. } & $1\times$ & $9.97\times$   & $79.6\times$ & $3608\times$ \\
    \hline
  \end{tabular}
  \vspace{-9pt}
  
\end{table}

\subsection{Performance and On-chip Memory Resource Analysis}
\subsubsection{Performance Analysis}
We use Algorithm~\ref{alg:spmm} for performance
analysis.
The dimension of three matrices 
$\mathbf{A}$, $\mathbf{B}$, and $\mathbf{C}$
are $M\times K$, $K\times N$, and $M\times N$
respectively. The number of non-zeros in sparse
matrix $\mathbf{A}$ is NNZ.
For the initialization of $\mathbf{C}$ (Line 2), 
$P$ PEs perform it in parallel, so the cycle
count is:
\vspace{-3pt}
\begin{equation} 
\label{eq:initC}
t_{\text{initC}}=K/P.
\vspace{-3pt}
\end{equation} 
At Line 4, a window of $\mathbf{B}$ is streamed in.
We partitioned the BRAM which stores $\mathbf{B}$
with a factor of $F_B=4$. The BRAM has two ports, so
we can store
$2\cdot F_B$ elements in one cycle. Thus
the cycle count for streaming in $\mathbf{B}$ is:
\vspace{-3pt}
\begin{equation} 
\label{eq:streamB}
t_{\text{streamB}}=K_0/(2\cdot F_B).
\vspace{-3pt}
\end{equation} 
In the PE region Line 6 - 10, the average
non-zeros for each $\mathbf{A}_{pj}$ is $\text{NNZ}/(P\times (K / K_0))$. So the
cycle count of Line 7 - 9 is:
\vspace{-3pt}
\begin{equation} 
\label{eq:cPE}
t_{\text{PE}}=(\text{NNZ}\times  K_0)/(P\times K).
\vspace{-3pt}
\end{equation} 
We process the element-wise computation of Line 13
with a parallel factor of $F_c\times N_0$, $F_c=16$. So
the cycle count is:
\vspace{-3pt}
\begin{equation} 
\label{eq:compC}
t_{\text{compC}}=M/F_C.
\vspace{-3pt}
\end{equation} 
The total cycle count is:
\vspace{-3pt}
\begin{equation} 
\label{eq:total}
\begin{split}
t
&=
(t_{\text{initC}}
+ (K/K_0)\times(t_{\text{streamB}} + t_{\text{PE}})
+ t_{\text{compC}})\times(N/N_0)\\
&=\left(
\frac{K}{2\cdot F_B} + 
\frac{\text{NNZ}}{P} +
\frac{M}{F_C} \right)\times\frac{N}{N_0}.
\end{split}
\vspace{-3pt}
\end{equation} 

\subsubsection{On-chip Memory Resource Analysis}

Storing $\mathbf{B}$ windows consumes BRAMs. 
One BRAM block is $1024\times18$ bits.
A window of $K_0=4096$ FP32 values requires
$4096/1024\times2=8$ BRAM blocks.
The partition factor is hidden because $F_B=4<8$.
With $N_0=8$ PUs for processing $N_0=8$ elements
of $\mathbf{B}$ in parallel, we assign $8\times N_0$
BRAM blocks for each PE.
Since a BRAM block has two ports, we share
one BRAM block between 2 PEs. 
So the total number of
BRAM blocks used is $8\times N_0\times P/2=2048$.

We use URAMs as much as possible.
A URAM block size is 4096$\times$72 bits. 
One URAM entry can store 2 FP32 values.
With $N_0=8$ PUs, we need 4 URAM blocks.
We set a URAM depth of 12288 for each PE.
So a total number of 
$12288/4096\times8/2\times=12\times64=768$ URAM blocks
are consumed, 
which accounts for 80\% of available URAMs
on a Xilinx U280 FPGA.

\section{Evaluation}
\label{sec:evaluation}
\subsection{Evaluation Setup}

\begin{table}[tb]
\caption{The specification of SpMM evaluation.}
\vspace{-9pt}
  \label{table:matrices}
  \centering
  \small
  \begin{tabular}{cc}
    \hline
    \textbf{Number of SpMMs}  & 1,400 \\
    \textbf{Number of Matrices}  & 200 \\
    \textbf{Row/column}  & 5 -- 513,351 \\
    \textbf{NNZ}  & 10 -- 37,464,962 \\
    \textbf{Density}  & 5.97E-6 -- 4.00E-1 \\
    \textbf{N}  & $N = 8, 16, 32, 64, 128, 256, 512$. \\
    \hline
  \end{tabular}
  \vspace{-6pt}
\end{table}

We evaluate on 1,400 SpMMs with 
200 sparse matrices and 7 N values ranging from 8 to 512.
Table \ref{table:matrices} illustrates
the properties of the sparse
matrices.
Of the 200 sparse matrices, 
we select
50 from SNAP
\cite{snapnets}
and 150 from SuiteSparse
\cite{davis2011university}.
We exclude matrices which are out-of-memory for a 5 GB memory budget.
The row/column number of the evaluated matrices
spans 5 to 513,351. The number of
non-zeros (NNZ) ranges from 10 to 
37,464,962. The density of the sparse matrices
ranges from 5.97E-6 to 4.00E-1.
The evaluated sparse matrices include
$73.5\%$ of all non-temporal matrices in SNAP
and \sextans supports $93.6 \%$ of matrices
in SuiteSparse. We evaluate single floating-point 
(FP32) SpMM.

\begin{table}[tb]
\caption{Process technology size, frequency, memory bandwidth, 
on-chip memory,
power, and
achieved peak SpMM
throughput of the four platforms.}
  \label{table:platforms}
  \vspace{-9pt}
  \centering
  \scriptsize
  \begin{tabular}{ccccccc}
    \hline
    & Tech. & Freq. & Bdw. & On-chip Mem. & Power & Peak Th.\\
    \hline
    \textbf{Tesla K80}    & 28 nm & 562 MHz   & 480 GB/s & 24.5MB & 130 W & 127.8 GFLOP/s\\
    \textbf{\sextansFPGA} & 16 nm & 189 MHz   & 460 GB/s & 22.7MB & 52 W & 181.1 GFLOP/s\\
    \textbf{Tesla V100}  & 12 nm & 1.297 GHz & 900 GB/s & 33.5MB & 287 W & 688.0 GFLOP/s\\
    \textbf{\sextansASIC}      & 16 nm & 350 MHz   & 900 GB/s & 24.5MB & 96 W & 343.6 GFLOP/s\\
    \hline
  \end{tabular}
  \vspace{-6pt}
  
\end{table}

\begin{table}[tb]
  \caption{Resource utilization of \sextansFPGA prototype on a Xilinx U280 FPGA board.}
  \label{table:FPGAultilization}
  \vspace{-9pt}
  \centering
  \small
  \begin{tabular}{cccc}
    \hline
    & Used & Available & Utilization (\%) \\
    \hline
    \textbf{BRAM}  & 3086    &   4032 &   76 \\
    \textbf{DSP48} & 3316    &   9024 &   36 \\
    \textbf{FF}    & 690,255 &   2,607,360 &   26 \\
    \textbf{LUT}   & 379,649 &   1,303,680 &   29 \\
    \textbf{URAM}  & 768     &   960 &   80 \\
    \hline
  \end{tabular}
  \vspace{-6pt}
\end{table}

\begin{figure}[tb]
\centering
\includegraphics[width=0.7\columnwidth]{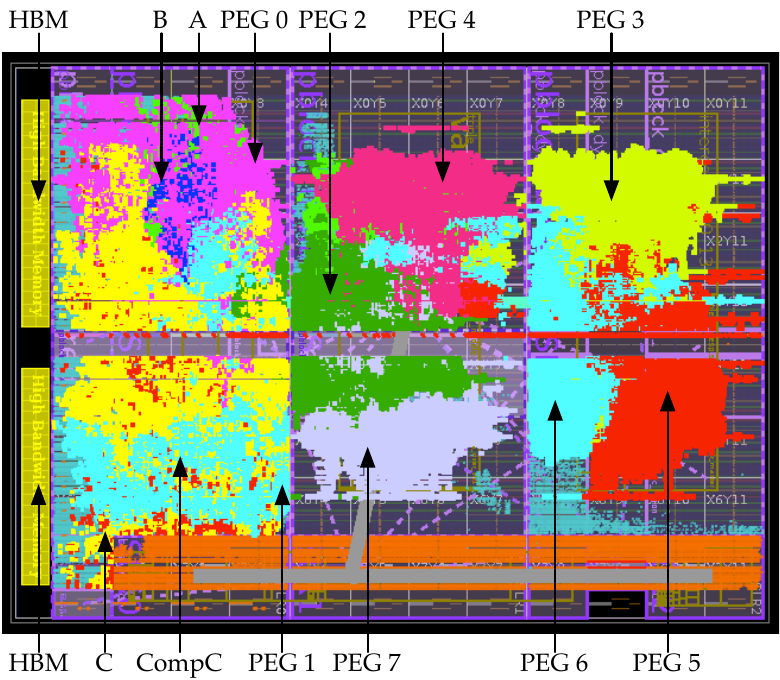}
\vspace{-9pt}
\caption{Layout of \sextansFPGA prototype on a U280 FPGA.}
\label{figure:layout}
\vspace{-15pt}
\end{figure}

\begin{figure*}[tb]
\centering
\includegraphics[width=2.05\columnwidth]{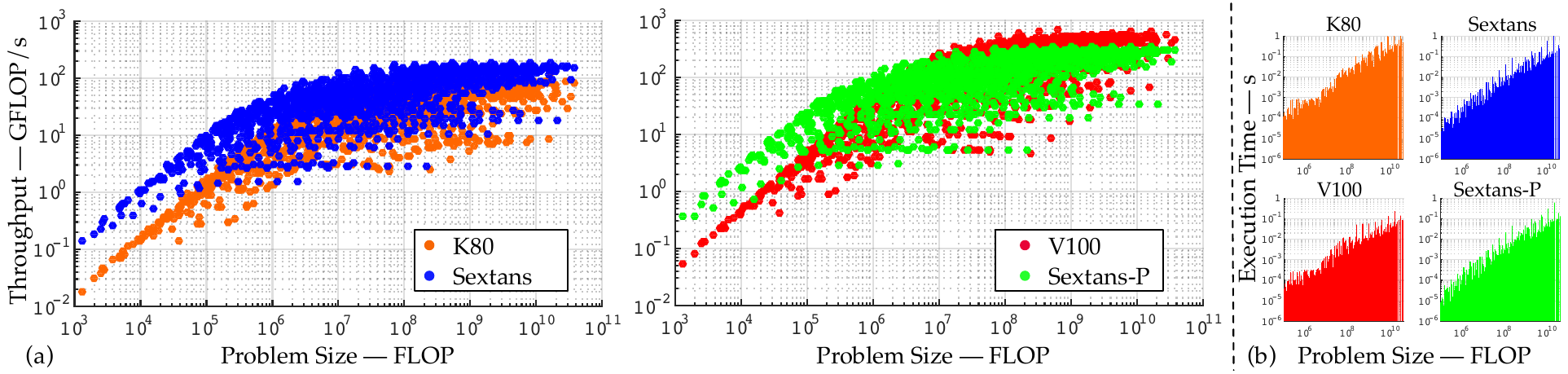}
\vspace{-9pt}
\caption{(a) Throughput (in GFLOP/s) and 
(b) execution time (in seconds) plotted with increasing problem size (in FLOP). The peak throughputs of K80, \sextansFPGA, V100, and \sextansASIC are 127.8 GFLOP/s, 
181.1 GFLOP/s, 688.0 GFLOP/s, and 343.6 GFLOP/s respectively. 
}
\label{figure:flops}
\vspace{-12pt}
\end{figure*}

We evaluate on four platforms -- an NVIDIA Tesla K80
GPU, an NVIDIA V100 GPU, an FPGA prototype 
\sextansFPGA and a projected prototype \sextansASIC with higher bandwidth
competitive to V100
and more frequency optimization.
The performance of K80, V100 and
\sextansFPGA is measured by runtime
and the performance of
\sextansASIC is simulated.
We list the specifications of the four evaluation
platforms in Table~\ref{table:platforms}.
For \sextans accelerator, we first prototype 
the accelerator on a Xilinx U280 HBM FPGA. 
The FPGA prototype provides both
a verification of the
\sextans architecture and a reference performance model
for
simulation. Moreover, \sextansFPGA is a highly
compatible working prototype which 
can be deployed in data center.
For a fair comparison,
we select two GPUs. K80 has a
memory bandwidth of 480 GB/s which is comparable
to the memory bandwidth of U280, i.e. 460 GB/s,
since memory accessing and
bandwidth are critical for sparse
and graph workloads \cite{ahn2015scalable,ham2016graphicionado,yang2018design}.
K80 is a more powerful platform than \sextansFPGA
because the frequency of K80 (562 MHz) is much higher than
the frequency of \sextansFPGA (189 MHz) and the memory
bandwidth of K80 is also slightly higher.
We measure the power consumption
of FPGA by Xilinx Board Utility {\tt xbutil} and the power of GPUs
by {\tt nvidai-smi}.
We also use a high-end GPU V100
and configure the memory bandwidth, \sextansASIC
according to that of V100 for fair comparison.
We set the frequency of
\sextansASIC to achieve
350 MHz
with the help of Autobridge~\cite{guo2021autobridge}
which is ongoing (most designs achieved
350 MHz with Autobridge).
According to $P=C\cdot V^2\cdot f$, we project the measured power of
\sextansFPGA by frequency
increase to 96 W as the
power of \sextansASIC.
Table~\ref{table:platforms} also lists 
the on-chip memory size which is sensitive to
memory-bound applications and the peak SpMM throughputs of the four platforms.

For the two GPU platforms, we use CuSPARSE \cite{naumov2010cusparse}
 routine {\tt csrmm} for the execution of floating-point SpMM.
The CUDA version is 10.2. We measure the GPU execution time
with CUDA runtime API
{\tt cudaEventElapsedTime} \cite{gputimer}. 
For the
FPGA prototype, we use Xilinx high level synthesis (HLS)
tool Vitis 2019.2. We list the
resource utilization of \sextansFPGA 
in Table \ref{table:FPGAultilization}.
The utilization rates of block RAM (BRAM)
and ultra RAM (URAM) are higher than other resources
because SpMM is memory intensive.
We show the layout of the \sextansFPGA accelerator
in Figure \ref{figure:layout}. We only
highlight the main components of the accelerator
including eight processing engine groups 
(denoted by PEG 0 -- 7 in Figure \ref{figure:layout}), memory reading units
for sparse matrix $\mathbf{A}$ and dense matrix 
$\mathbf{B}$ (denoted by A, B in Figure \ref{figure:layout}),
memory reading and writing units
for dense matrix 
$\mathbf{C}$ (denoted by C in Figure \ref{figure:layout}),
and the compute units for partial $\mathbf{C}$
(denoted by CompC in Figure \ref{figure:layout}).
We launch the FPGA accelerator with OpenCL \cite{stone2010opencl} runtime
and measure the FPGA execution time.
We build an in-house simulator
to simulate the performance 
of \sextansASIC after the prototyping on
FPGA. The simulator is based on the emulation
C++ code of \sextans.
Since \sextans is a streaming accelerator,
we model the computing time and memory accessing
time and record the larger one as the processing
time at each stage. The frequency and memory bandwidth
in the simulation is configured the same as
V100 that we list in Table \ref{table:platforms}. 

\subsection{Results}

\subsubsection{Overall Performance.}

\begin{figure*}[tb]
\centering
\includegraphics[width=1.5\columnwidth]{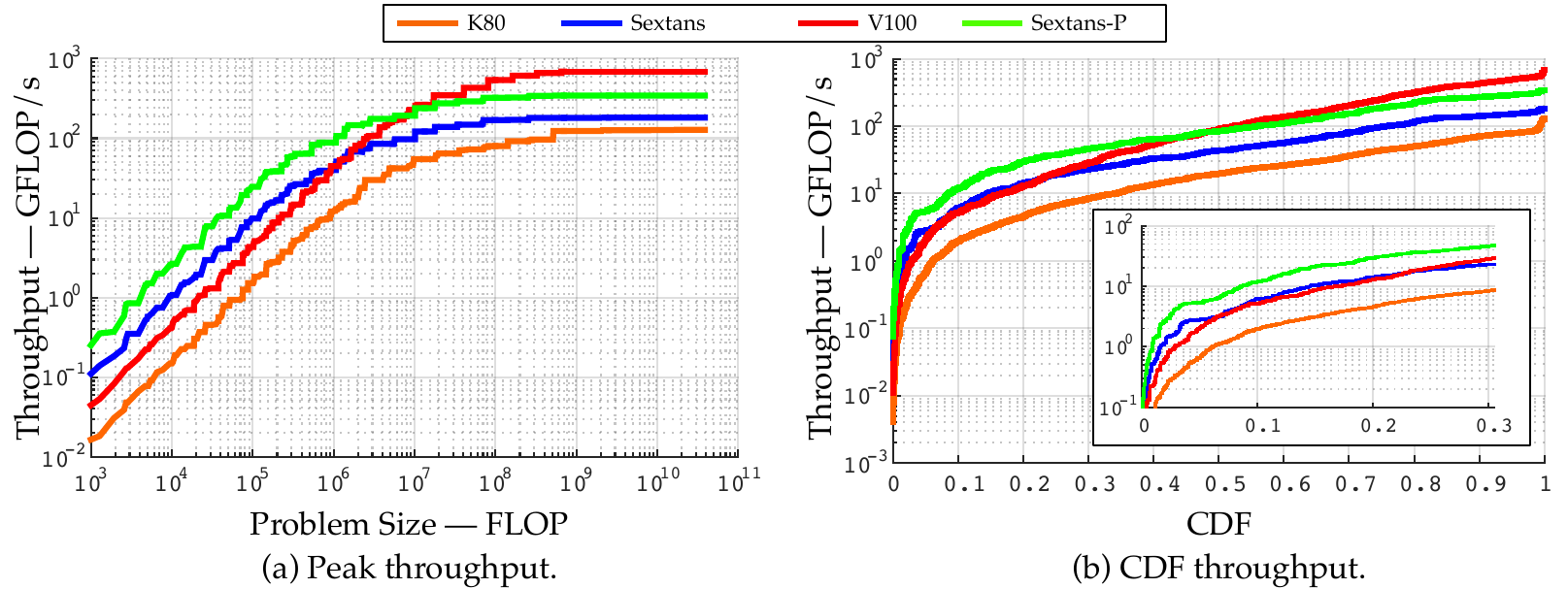}
\vspace{-9pt}
\caption{(a) Peak throughput with the increase of problem size and (b) CDF throughput of the four evaluated platforms.}
\label{figure:peak_cdf}
\vspace{-6pt}
\end{figure*}

We plot the throughput (in GFLOP/s)
of 2,000 SpMMs
of 200 sparse matrices with 7
N configurations (N = 8, 16, 32, ..., 512) on the four platforms
in Figure \ref{figure:flops} (a)
and the execution time in second
in Figure \ref{figure:flops} (b)
with the increase of the problem size
in FLOP.
The problem size is defined as
the number of floating-point 
operations for executing one
SpMM $\mathbf{C} = \alpha\cdot\mathbf{A}\times\mathbf{B} + \beta\mathbf{C}$, and 
the problem size is proportional to N.
The throughput is calculated as $p/t$ 
where $p$ is the problem size and $t$  is the execution time.
Overall, the peak throughputs of K80, \sextansFPGA, V100, and \sextansASIC are 127.8 GFLOP/s, 
181.1 GFLOP/s, 688.0 GFLOP/s, and 343.6 GFLOP/s respectively. The geomean speedups of the four
platforms normalized to K80 are 1.00$\times$,
2.50$\times$, 4.32$\times$, and 4.94$\times$ respectively. The 
geomean speedup of \sextansASIC to
V100 is 1.14$\times$.

From Figure \ref{figure:flops} (a)
we see the overall trend that
with the increase of the problem size
the throughput of the four platforms
increases and after a problem size
threshold is reached,
the throughput gets saturated
at the peak throughput of the four 
platforms.
We compare \sextansFPGA with K80 in one
subfigure and \sextansASIC with V100
in another because the two platfroms 
in the same subfigure are comparable.
We see the throughput
dots of \sextansFPGA at the top of
dots of K80.
The dots of \sextansASIC are higher than
the dots of V100 for problem size $<10^7$
FLOP.
Although \sextansASIC has the same memory bandwidth 
as V100, the frequency of V100 is much higher than 
that of \sextansASIC (1297 MHz v.s. 350 MHz). So
the saturated throughput of V100 is higher than
that of \sextansASIC.

For a problem size less than $10^6$ FLOP,
we see the throughput increases with the
increase of problem size for the four platforms,
because there are setup and ending processing overheads
for the four platforms. For example, in \sextans architectures,
on-chip memory for partial $\mathbf{C}$ is initialized
before the main processing 
loop and $\mathbf{C}$ is written back
to off-chip memory after the main processing loop. 
GPUs also need similar setup processing and writing
back from on-chip buffer to device memory.
However, with the increase of the problem size the
overhead is amortized and it is better to parallelize
a larger size problem. So we see the performance 
(GFLOP/s) increases.
Notice that for problem size less than
$10^6$ FLOP, \sextansFPGA performs better
than both K80 (computing power comparable with \sextansFPGA) and V100 (computing power much higher than \sextansFPGA).
That is because CUDA runtime launches
GPU SpMM kernels. The CUDA runtime has a small
overhead which is not observable on large-size
problems but on problems less than $10^6$ FLOP,
the overhead degrades the GPU performance. 
However, FPGA accelerators can fuse two or
more kernels into one so the runtime overhead
between kernels can be eliminated.


From Figure \ref{figure:flops} (b)
we see the execution time
decreases successively from
K80, \sextansFPGA, V100 to
\sextansASIC. For each specific 
platform, the execution time increases
with the increase of the problem
size. We can also see spikes
in the execution time plots
and dots deviating from the throughput trend.
The trend of throughput and execution time can be
mainly determined by the problem size for a specific 
platform, but for a specific sparse matrix,
the non-zero distribution pattern also determines
the execution time. So for matrices with a close
problem size, the various non-zero distributions
lead to distinguished execution times, which are
reflected as spikes in Figure \ref{figure:flops} (b)
and deviating dots in Figure \ref{figure:flops} (a).

\subsubsection{Peak and CDF Performance.}

To better understand the performance
of K80, \sextansFPGA, V100, and 
\sextansASIC, we show the peak throughput
with the increase of the problem size
in Figure \ref{figure:peak_cdf} (a)
and the cumulative distribution function
(CDF) throughput
in Figure \ref{figure:peak_cdf} (b).
The peak throughput is defined as
the maximum throughput of all
problems whose size is smaller than
a specific problem size (X-axis).
The peak plot helps us to understand the peak
performance on problems less than a specific size
and the CDF plot helps to capture every performance
point.

We compare the problem size for
a specific platform to reach the
peak throughput.
\sextansFPGA and \sextansASIC is the most efficient
because \sextansFPGA and \sextansASIC reaches its peak
throughput at a smallest problem size
around $8\times10^7$ FLOP. 
For the two GPU platforms they
reach their peak throughput around
$\times10^9$ FLOP. 
In Figure \ref{figure:peak_cdf} (b)
we see \sextansASIC has the highest
throughput compared to
the other platforms
for CDF < 0.5.
The larger gap indicates higher 
efficiency of one platform compared 
with another. 
We observe that for a problem size
less than $10^6$ FLOP and CDF
less than 0.1, the throughputs of
\sextansFPGA is higher than both
K80 (computing power comparable with \sextansFPGA) and V100 (computing power much higher than \sextansFPGA).
As we discussed before, that is because the 
small CUDA runtime is amplified on small-size problems.

\subsubsection{Memory Bandwidth Utilization.}

\begin{figure}[tb]
\centering
\vspace{0pt}
\includegraphics[width=0.75\columnwidth]{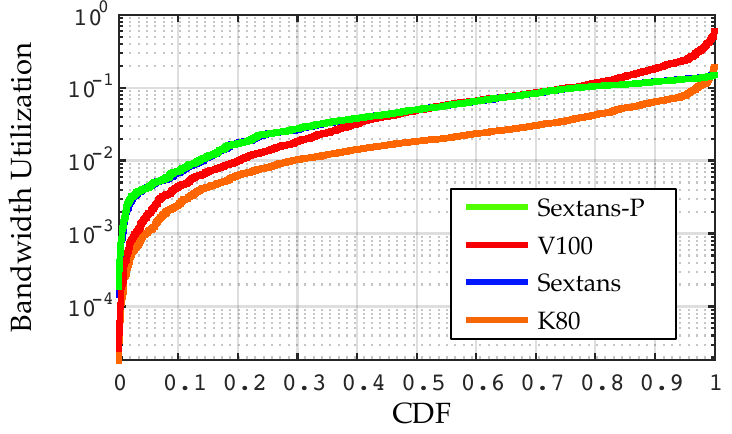}
\vspace{-9pt}
\caption{Memory bandwidth utilization.}
\label{figure:bandwidth}
\vspace{-15pt}
\end{figure}

We compare the memory bandwidth utilization
of the four evaluated platforms in
Figure \ref{figure:bandwidth}. 
The memory bandwidth utilization is defined as
$(4\times(\text{NNZ} + N\times(2\times M + K)))/t/\text{Bdw}$
where NNZ is the number of 
non-zeros of a sparse
matrix, M is the number of 
rows of sparse matrix $\mathbf{A}$ and
also the number of rows
of dense matrix $\mathbf{C}$,
N is the number of columns
of dense matrix $\mathbf{B}$ and $\mathbf{C}$, and Bdw is 
the maximum available 
memory bandwidth as listed
in Table \ref{table:platforms}. 
A factor 4 is multiplied 
because a floating-point value occupies 4 bytes. 
A factor 2 is multiplied
to M because matrix $\mathbf{C}$ is read 
and written once each.
Notice that the memory bandwidth utilization
is not a memory bandwidth occupation rate which is calculated by (Occupied Bandwidth)/(Maximum Available Bandwidth).
We do not use memory bandwidth occupation rate because
an inefficient design where memory bandwidth is fully occupied
but nothing is done can achieve a $100\%$ memory bandwidth occupation rate. A higher memory bandwidth utilization
is better but one cannot achieve
a $100\%$ memory bandwidth utilization
for two reasons. 
First, it is impossible
to read/write all matrices once which 
requires the on-chip
memory to be as large as the off-chip memory.
Second, for sparse matrix we only count NNZ here but 
the index also occupies memory space. For example, besides the 4 bytes
for the value of a non-zero,
coordinate list (COO) format uses another 4 bytes for
row index and 4 bytes for column index. In compressed
sparse row (CSR) format, another 4 bytes are needed by the index
of each non-zero while the row index is compressed and 
the compressed row also occupies extra memory.

The geomean bandwidth utilization of
the four platforms are
$1.47\%$,
$3.85\%$, 
$3.39\%$, and 
$3.88\%$ respectively.
Memory bandwidth utilization is 
relatively low
for sparse workloads \cite{ahn2015scalable,ham2016graphicionado,yang2018design}
especially for small size sparse matrices. 
SpMM is memory bound and the memory bandwidth utilization
of \sextansASIC is $1.15\times$ of V100. \sextansASIC and
V100 work at 
the same memory bandwidth (900 GB/s).
The memory bandwidth utilization translates to the 
$1.14\times$ geaomean speedup of \sextansASIC
compared with V100.
\sextansFPGA
achieves a $2.62\times$ bandwidth utilization
compared to K80. The frequency and memory
bandwidth of K80 is higher than 
that of \sextansFPGA
(562 MHz / 480 GB/s v.s. 189 MHz / 460 GB/s).
So \sextansFPGA achieves a $1.68\times$
geomean speedup compared to K80.
Although the memory bandwidth utilization
of \sextansFPGA is $1.14\times$ compared to
the memory bandwidth utilization of V100,
the frequency and memory
bandwidth of \sextansFPGA is much
lower than that of V100
(189 MHz / 460 GB/s v.s. 1297 MHz / 900 GB/s).
So the geomean speedup of \sextansFPGA
is lower than that of V100.


\begin{table*}[tb]
\caption{Comparison with related sparse and dense accelerators.}
  \label{table:comparison}
  \footnotesize
  \centering
  \vspace{-9pt}
  \begin{threeparttable}
  \begin{tabular}{ccccccccc}
    \hline
    & \textbf{Kernels} & \textbf{Mat. NNZ} & \textbf{Prob. Size} &
    \textbf{Throughput} & \textbf{FPGA} & \textbf{Simulation} & \textbf{Real Exe.}& \textbf{HFlex}\\
    \hline
    T2S-Tensor \cite{srivastava2019t2s} & Dense MM, MV, etc\tnote{1} & $2\times10^3$ & - &
    738 GFLOP/s & \textbf{Yes} & No & \textbf{Yes} & No \\
    
    AutoSA \cite{wang2021autosa} & Dense MM, etc\tnote{1} & $4\times10^6$ & $7\times10^9$ &
    950 GFLOP/s & \textbf{Yes} & No & \textbf{Yes} & No \\
    
    Tensaurus \cite{srivastava2020tensaurus} & SpMV, SpMM, etc\tnote{2} & $4.2\times10^6$ & - &
    512 GFLOP/s\tnote{3} & No & \textbf{Yes} & No & No \\

    \hline
    \cite{fowers2014high} & SpMV & $5\times10^6$ & $<1\times10^7$ &
    3.9 GFLOP/s & \textbf{Yes} & No & \textbf{Yes} & No \\

    Spaghetti \cite{hojabr2021spaghetti} & SpGEMM & $1.6\times10^7$ & - &
    27 GFLOP/s & \textbf{Yes} & No & \textbf{Yes} & No \\

    ExTensor \cite{hegde2019extensor} & SpMM, SpGEMM, etc\tnote{2} & $6\times10^6$ & - &
    64 GFLOP/s & No & \textbf{Yes} & No & No \\
    
    SIGMA \cite{qin2020sigma} & SpGEMM & - & - &
    - & No & \textbf{Yes} & No & No \\

    SpArch \cite{zhang2020sparch} & SpGEMM & $1.65\times10^7$ & - &
    10.4 GFLOP/s & No & \textbf{Yes} & No & No \\

    OuterSPACE \cite{pal2018outerspace} & SpGEMM & $1.65\times10^7$ & - &
    2.9 GFLOP/s & No & \textbf{Yes} & No & No \\

    SpaceA \cite{xie2021spacea} & SpMV & $1.4\times10^7$ & $1.43\times10^7$ &
    - & No & \textbf{Yes} & No & No \\
    
    \sextansFPGA & SpMM & $3.7\times10^7$ & $3\times10^{10}$ &
    181.1 GFLOP/s & \textbf{Yes} & No & \textbf{Yes} & \textbf{Yes} \\
    
    \sextansASIC & SpMM & $3.7\times10^7$ & $3\times10^{10}$ &
    343.6 GFLOP/s & No & \textbf{Yes} & No & \textbf{Yes} \\
    \hline
  \end{tabular}

  \begin{tablenotes}
        \footnotesize
        \item[1] Other dense tensor kernels such as TTMc, MTTKRP are also supported. $^2$ Other dense tensor kernels such as sparse TTM, sparse TTV are also supported.
        \item[3] 512 GFLOP/s is achieved on dense multiplication, and the throughput of sparse multiplication is lower.  
      \end{tablenotes}
    \end{threeparttable}
    \vspace{-12pt}
  
\end{table*}

The maximum memory bandwidth utilization of
K80, \sextansFPGA, V100, and \sextansASIC 
are 
$19.00\%$,
$14.92\%$,
$59.96\%$, and
$14.96\%$ respectively.
V100 achieves the highest memory bandwidth utilization
and is significantly better than the other three platforms
according to Figure \ref{figure:bandwidth}.
The maximum memory bandwidth utilization of 
FPGAs is lower than that of GPUs
because of HLS tool limitation. 
The Xilinx HLS tool requires users to handle
the connection of memory pointers to M AXI bundles. 
For SpMM we need to handle memory pointers
of  matrix $\mathbf{A}$, $\mathbf{B}$,  $\mathbf{C}$, and pointer list $\mathbf{Q}$.
The Xilinx U280 platform allows a maximum number of 32 M AXIs. There are 32 
pseudo HBM channels. One memory pointer
can only be mapped to one M AXI. 
Thus, the number of parallel HBM channels
is significantly limited. As a result, 
the memory bandwidth utilization is low. 

\subsubsection{Energy Efficiency.}

\begin{figure}[tb]
\centering
\includegraphics[width=0.8\columnwidth]{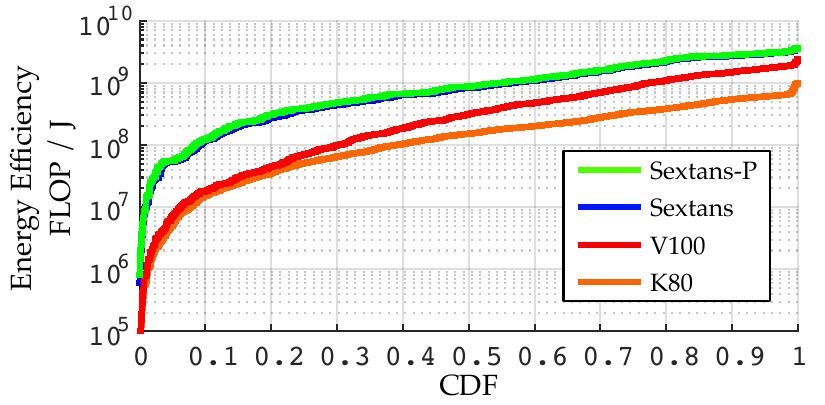}
\vspace{-9pt}
\caption{Energy efficiency of the four evaluated platforms.}
\label{figure:energy}
\vspace{-15pt}
\end{figure}

We compare the energy efficiency
of the four evaluated platforms in
Figure \ref{figure:energy}. 
The energy
efficiency is defined as $p/(t\cdot\text{Power})$ where
$p$ is the problem size,
$t$ is the execution time,
and Power is the power consumption of a specific platform
(listed in Table \ref{table:platforms}).
The energy efficiency of \sextansFPGA
and \sextansASIC is similar.
The geomean energy efficiency of
K80, \sextansFPGA, V100, and \sextansASIC 
 are
$1.06\times10^8$ FLOP/J,
$6.63\times10^8$ FLOP/J,
$2.07\times10^8$ FLOP/J, and 
$7.10\times10^8$ FLOP/J respectively.
\sextansFPGA is $6.25\times$ better than
K80 and $3.20\times$ better than
V100. 
Normalized to
K80, the energy efficiency improvements
of \sextansFPGA, V100, and \sextansASIC
are $6.25\times$,
$1.95\times$, and
$6.70\times$, respectively.
The maximum energy efficiency of
the four platforms
are 
$9.83\times10^8$ FLOP/J,
$3.48\times10^9$ FLOP/J,
$2.40\times10^9$ FLOP/J, and
$3.60\times10^9$ FLOP/J respectively.

\subsection{Comparison with Related Accelerators}

We compare \sextansFPGA and \sextansASIC with accelerators for similar
sparse workloads in Table \ref{table:comparison}.
We also include the accelerators for dense workloads
as a comparison to see the performance gap between
sparse and dense accelerators.
The related accelerators include
FPGA implementation and 
ASIC simulation works and the supported
workloads include dense matrix matrix/vector multiplication
(MM, MV), SpMM, sparse matrix vector
multiplication (SpMV) and sparse matrix - sparse matrix multiplication
(SpGEMM). Some accelerators also
target for high order tensor
operation such as 
metricized tensor times Khatri Rao product (MTTKRP) and tensor times matrix chain (TTMc).
Compared with the related works,
\sextans supports the 
largest sparse problem size
(largest matrix NNZ and problem size) and achieves the highest
throughput.
\sextansFPGA is the only real executable
accelerator for supporting SpMM.
\sextans is the only accelerator which features
the hardware flexibility (HFLex)
which supports arbitrary problems directly
by the hardware.
For existing accelerators, to
support a different problem 
configuration, the accelerator 
architecture must be modified.
For ASIC accelerators, 
an architecture re-design and
chip tape-out can take anywhere from weeks to months.
For FPGA accelerators even with
an automatic design flow such as
AutoSA \cite{wang2021autosa}
where the architecture re-design
time is saved,
a new design will still need
many hours or even a few days
due to long synthesis and 
place/route time.
Although an accelerator for
a fixed-size problem 
can be built
and
a new problem can be decomposed
as multiple kernels for
the fixed-size accelerator,
the runtime overhead
for launching multiple kernel 
is high (0.15ms$\times$ launching times).
The HFlex feature enables
\sextans to support a new
problem without re-running
the time-consuming flows
including synthesis/place/route
for both FPGA and ASIC and a fabrication
for ASIC.


\section{Conclusion}
\label{sec:conclusion}
SpMM is the key operator for a wide range of applications. 
We present \sextans, an accelerator for general-purpose SpMM processing. 
We propose (1) HFlex processing to enable 
prototyping the hardware accelerator once
to support all SpMMs as a general-purpose
accelerator, (2) PE-aware non-zero scheduling
for balance workloads with an II=1 pipeline, and
(3) on-chip and off-chip memory
optimization to resolve the challenge of
inefficient random memory accessing and off-chip 
accessing of large matrices.
We present
an FPGA prototype Sextans which is executable on a Xilinx U280 HBM FPGA board and a projected prototype Sextans-P with higher bandwidth competitive to V100 and more frequency optimization. We conduct comprehensive evaluation on 1,400 SpMMs on 200 matrices
to compare Sextans with NVIDIA K80 and V100 GPUs. Sextans achieves a 2.50x geomean speedup over K80 GPU and Sextans-P achieves a 1.14x geomean speedup over V100 GPU (4.94x over K80).

\vspace{6pt}
\noindent{\bf \large ACKNOWLEDGMENT}

We thank the anonymous reviewers and our labmates Weikang Qiao, Zhe Chen, and Licheng Guo, for their valuable feedbacks.
This work is supported in part
by NSF RTML Program (CCF-1937599),
CDSC industrial partners~\footnote{\url{https://cdsc.ucla.edu/partners}}
and Xilinx XACC Program.

\bibliographystyle{ACM-Reference-Format}
\bibliography{references}

\end{document}